\documentclass[twocolumn,showpacs,prl]{revtex4}
\usepackage{amsmath}
\usepackage{epsf}
\usepackage{graphicx}
\usepackage{dcolumn}
\usepackage{bm}

\begin{document}

\title{Thermodynamics  of a Fermi liquid beyond the
 low-energy limit}
\author{Andrey V. Chubukov$^{1},$ Dmitrii L. Maslov$^{2},$ Suhas Gangadharaiah$^{2}$%
, and Leonid I. Glazman$^{3}$}
\date{\today}

\begin{abstract}
We consider the non-analytic temperature dependences of the specific heat
coefficient, $C(T)/T$, and spin susceptibility, $\chi _{s}\left( T\right) ,$
of 2D interacting fermions beyond the weak-coupling limit. We demonstrate
within the Luttinger-Ward formalism that the leading temperature dependences
of $C(T)/T$ and $\chi _{s}(T)$ are linear in $T$, and are described by the
Fermi liquid theory. We show that these temperature dependences are
universally determined by the states near the Fermi level and, for a generic
interaction, are expressed via the spin and charge components of the exact
backscattering amplitude of quasi-particles. We compare our theory to recent
experiments on monolayers of He$^{3}$.
\end{abstract}

\affiliation{ 
$^1$Department of Physics, University of Maryland,
College Park, MD 20742-4111\\
$^{2}$Department of
Physics, University of Florida, P. O. Box 118440, Gainesville, FL
32611-8440 \\
$^3$Theoretical Physics Institute, University of Minnesota,
Minneapolis, MN 55455}
\maketitle

The Landau Fermi Liquid (FL) theory states that the low-energy properties of
an interacting fermionic system are determined by the states in the vicinity
of the Fermi surface, and are similar to that of weakly interacting
quasi-particles. At the lowest temperatures, when decay of quasi-particles
can be neglected, the specific heat $C(T)\propto T$ and spin susceptibility $%
\chi _{s}(T)=$const of a FL differ from the corresponding quantities for the
Fermi gas only via the renormalizations of the effective mass and $g-$
factor \cite{agd}. However, this low-temperature limit of the FL theory,
considered by Landau,
 cannot tell whether the subleading terms in $T$ are analytic, and
whether they come only from low-energy states (and are therefore described
by the FL theory) or from the states far away from the Fermi surface.

For non-interacting fermions, the subleading terms in $C(T)/T$ and $\chi
_{s}(T)$ scale as $T^{2}$ and come from high-energy states. However, it was
found back in the 60s that in 3D systems, the leading correction to $C(T)/T$
due to interaction with either phonons~\cite{eliashberg} or paramagnons~\cite
{doniach} is non-analytic in $T$ ( $T^{2}\ln T$)  and comes from the states
in the immediate vicinity of the Fermi surface. The same result was later
shown to hold for the electron-electron interaction \cite{amit}. More
recently, it was shown by various groups~\cite
{bedell,belitz,dassarma,millis,marenko,chm} that the temperature
dependence of $C(T)/T$ is also non-analytic in 2D and starts with a linear-in-$T$
term. The same behavior was also found for the uniform spin
susceptibility~\cite{belitz,marenko,millis,chm}. Two of us have argued~\cite
{chm} that, to second order in short-range interaction, these linear-in-$T$
terms originate exclusively from scattering of fermions with zero total
momentum and either small or near $2k_{F}$ momentum transfers (
``backscattering'').

In this paper, we consider the specific heat and spin susceptibility for a
generic 2D Fermi liquid. In this case~\cite{agd}, the leading (constant)
terms in $C(T)/T$ and $\chi _{s}(T)$ are expressed via the two harmonics --$%
F _{c}^{(1)}$ and $F _{s}^{(0)}$-- of the quasiparticle
interaction function $F (\theta )$, or, equivalently, via the same
harmonics of the scattering amplitude $A(\theta )$ ($c$ and $s$
refer to charge and spin components). We show that the
linear-in-$T$ terms in $C(T)/T$ and $\chi _{s}(T)$ are also
universally expressed via the scattering amplitude, but they are
determined by $A(\theta )$ at a \emph{particular angle} $\theta
=\pi ,$ rather than by $A(\theta )$ averaged over the Fermi
surface. As there is no simple relation between $A(\pi )$ and $F (\pi )$%
, these subleading terms cannot be simply expressed via the Landau
function. The only exception is a weakly screened Coulomb
interaction, when the $T$-term in $C(T)/T$ can be expressed via $F
(\pi )$.

\begin{figure}[tbp]
\begin{center}
\epsfxsize=0.8\columnwidth \epsffile{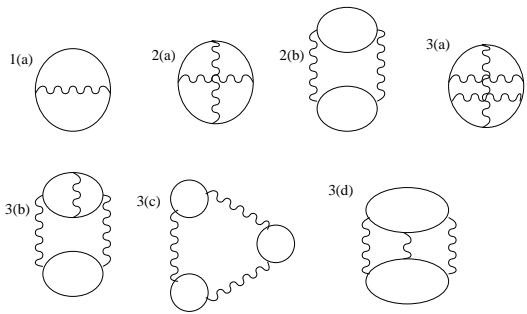}
\end{center}
\caption{Non-trivial second-order and third-order Diagrams for the
thermodynamic potential. For the Coulomb potential, diagrams 1(a), 2(b) and
3(c) represent ring series.}
\label{fig:1}
\end{figure}

To shorten the presentation, we discuss in some detail
the calculation for $C(T)$,
 and then just present the result for $\chi _{s}(T),$ which can
obtained in a similar manner~\cite{chi_s}. The most straightforward way to obtain $C(T)$ beyond
the leading term in $T$ is to find the thermodynamic potential $\Xi (T)$
within the Luttinger-Ward approach~\cite{luttinger},  and use $%
C(T)=-T\partial ^{2}\Xi /\partial ^{2}T$. The non-analyticity of $C(T)/T$
obviously originates from the non-analyticity of $\Xi (T)$. The
thermodynamic potential $\Xi $ is expressed as
\begin{equation}
\Xi =\Xi _{0}-2T\sum_{\omega _{m}}\int \frac{d^{2}k}{4\pi ^{2}}\left[ \ln
[G_0 G^{-1}]-\Sigma G+\sum_{\nu }\frac{1}{2\nu }\Sigma _{\nu }G\right] ,
\label{o1}
\end{equation}
where $\Xi _{0}$ is the thermodynamic potential of the free Fermi gas per
unit area, $G_{0}=\left( i\omega _{m}-\epsilon _{k}\right) ^{-1}$, $G=\left(
i\omega _{m}-\epsilon _{k}+\Sigma \right) ^{-1}$, $\Sigma $ is the exact (to
all orders in the interaction) self-energy, and $\Sigma _{\nu }$ is the
skeleton self-energy of order $\nu $. Both $\Sigma _{\nu }$ and 
the sum $\Sigma =\sum_{\nu }\Sigma _{\nu }$ are evaluated at
\textit{finite} $T$.
Diagrams associated with the first two terms in (\ref{o1}) correspond to the
self-energy insertions into the free thermodynamic potential $\Xi
_{0}=2T\sum_{\omega _{m}}\int d^{2}k/\left( 2\pi \right) ^{2}\ln {G_{0}(}%
\omega _{m},k{)}$, diagrammatically represented by a loop (Fig.~\ref{fig:1},%
1a). One can readily verify that such diagrams simply renormalize the constant
term in $C(T)/T$. The non-analytic temperature dependence of $C(T)/T$ comes
from the third
 term in (\ref{o1}).


To understand the origin of the non-analyticity in $\Xi (T)$, consider first
a weak short-range interaction $U(q)$. To second order in $U$, the skeleton
term gives rise to diagrams 2a and 2b in Fig.~\ref{fig:1}.
 Assume momentarily that $U$ is a constant. Then each of the two diagrams
can be re-expressed as a product of two particle-hole bubbles $\Pi (q,\Omega
_{m})$, so that
\begin{equation}
\delta \Xi =-\frac{1}{2}TU^{2}\sum_{\Omega }\int \frac{d^{2}q}{4\pi ^{2}}\Pi
^{2}(q,\Omega _{m}),  \label{o_1}
\end{equation}
where $\delta \Xi \equiv \Xi -\Xi _{0}$.

It is intuitively clear that the non-analyticity in $\delta \Xi $ should be
related to a non-analyticity in $\Pi (q,\Omega _{m})$. There are two regions
of $q$ where $\Pi $ is non-analytic. First region is near $q=0$, where $\Pi
(q,\Omega _{m})=-(m/2\pi )(1-|\Omega _{m}|/(\Omega
_{m}^{2}+(v_{F}q)^{2})^{1/2})$. For $v_{F}q\gg |\Omega _{m}|$, the
Landau-damping term $\left( |\Omega _{m}|/q\right) $ is non-analytic in $q$.
This non-analyticity leads to a long-range tail of $\Pi (r,\Omega _{m})$ in
real space: $\Pi (r,\Omega _{m})\propto |\Omega _{m}|/r$. The second region
is near $2k_{F}$, where $\Pi (q,\Omega _{m})=-(m/2\pi )\left( 1-\left( \bar{q%
}+\sqrt{\bar{q}^{2}+\bar{\Omega}_{m}^{2}}\right) ^{1/2}\right) ,$ with $\bar{%
q}=(q-2k_{F})/2k_{F}~$and ${\bar{\Omega}}=\Omega /2k_{F}v_{F}$. The
singularity at $\bar{q}=0$ is known as the Kohn anomaly. Most of the effects
caused by the Kohn anomaly -- Friedel oscillations, Kohn-Luttinger pairing
-- are associated with the (asymmetric) singularity in the static bubble: $%
\Pi (\bar{q},\Omega _{m}=0)-\Pi (0,0)\propto \theta \left( \bar{q}\right)
\sqrt{\bar{q}}$. However, the static Kohn anomaly does not cause a
non-analytic $T$-dependence of $\delta \Xi (T)$ in (\ref{o_1}%
) as the momentum integral over static $\Pi (q,0)$ is infrared
convergent. The non-analyticity in $\delta \Xi (T)$ comes from the \textit{%
dynamic} Kohn anomaly, which is a term $|{\bar{\Omega}}|/\sqrt{|{\bar{q}}|}$
in $\Pi (\bar{q},\Omega _{m}=0)$ for $-\bar{q}\gg \left| \Omega _{m}\right|
/v_{F}$. This term leads to a \emph{dynamic} Friedel oscillation: $\Pi
(r,\Omega _{m})\propto |\Omega _{m}|\cos (2k_{F}r)/\sqrt{r}$.

Integrating in Eq. (\ref{o_1}) over the two momentum regions where $\Pi
\left( q,\Omega _{m}\right) $ is non-analytic, we find that each of these
two regions contributes a logarithmic singularity of the form $\Omega
_{m}^{2}\ln |\Omega _{m}|$, the prefactors
 are  the same. This logarithmic
singularity is the key effect. Had it been absent, the Matsubara sum of $%
\Omega _{m}^{2}+\Omega _{m}^{4}+\dots $ would have been controlled by
high frequencies, of order $E_{F},$ and would have led to an analytic
expansion $\delta \Xi =$const$+T^{2}+T^{4}+...$. The presence of the
logarithm changes the story, as now the frequency sum contains a universal
contribution from frequencies of order $T$. Using the Euler-Maclaurin
summation formula, we obtain
\begin{eqnarray}
&&\delta \Xi =-\frac{4\pi u^{2}T^{3}}{v_{F}^{2}}~S\left( M\right) ,~~S\left(
M\right) \equiv \sum_{m=0}^{M}m^{2}\ln \frac{M}{m}  \notag \\
&=&\frac{1}{9}M^{3}-\frac{1}{12}M-\frac{\zeta \left( {3}\right) }{4\pi ^{2}}+%
\frac{1}{360M}+\dots   \label{e2}
\end{eqnarray}
where $u=mU/2\pi $, and $M\sim E_{F}/2\pi T\gg 1$. Those terms in (\ref{e2})
that depend on $M$ yield a regular expansion for $\delta \Xi $ in powers of $%
T^{2}$, whereas the $M$-independent term -- the third term in the second
line of Eq.(\ref{e2}) -- gives rise to
 $\delta \Xi
\propto T^{3}$, and hence to a $T^{2}-$ term in $C(T)$.

Next, we take a more careful look at which four-fermion vertices actually
contribute to the non-analytic part of $C(T)$. For the $2k_{F}$ -part, the
answer follows immediately from the observation that, for a given
direction of $\mathbf{q}$, the Kohn anomaly (both static \textit{and}
dynamic) comes from the internal fermionic momenta near $\mathbf{q}/2$ and $-%
\mathbf{q}/2$ in both bubbles in diagrams  2a and 2b
of Fig.\ref{fig:1}. The relevant vertex then
has the momentum structure $(\mathbf{k,-k;-k,k})$, which corresponds to
backscattering.

For the $q=0$ part, the momentum structure is less obvious as, at the first
glance, the internal momenta in the two bubbles in diagrams
 2a and 2b are uncorrelated. However, the logarithmic singularity of the momentum integral
comes only from the $|\Omega _{m}|/q-$ part of each polarization bubble. One
can show that this part comes from the integration over that special region
where the internal momenta are nearly orthogonal to $\mathbf{q}$. Since the
relevant momenta in the two bubbles are almost orthogonal to the same vector
($\mathbf{q)}$, they must be either nearly parallel or nearly antiparallel
to each other. We verified that the contribution from the near parallel
momenta, \emph{i.e.}, from forward scattering, vanishes and the full result
comes from nearly antiparallel momenta. This implies that the $q=0$
contribution to $\delta \Xi $ involves a vertex with the momentum structure $%
(\mathbf{k,-k;k,-k})$. This vertex is also a part of the backscattering
 amplitude.

We can now extend our second-order analysis to a finite-range interaction $%
U(q)$. That only backscattering is relevant means that only $U(0)$ and $%
U(2k_{F})$ contribute to the $T$ -term in $C(T)/T$. The contribution of
the diagram 2b is proportional to $U^{2}(0)+U^{2}(2k_{F})$,
whereas that of the diagram 2a is proportional to $U(0)U(2k_{F})$.
Collecting the prefactors, we obtain for the non-analytic part of the
specific heat $\Delta C\equiv C(T)-\gamma T$
\begin{equation}
\Delta C\left( T\right) /T=-\left(
u_{0}^{2}+u_{2k_{F}}^{2}-u_{0}u_{2k_{F}}\right) ~\frac{3m\zeta (3)}{\pi }~%
\frac{T}{E_{F}},  \label{genu}
\end{equation}
where $u_{0}=mU(0)/2\pi $ and $u_{2k_{F}}=mU(2k_{F})/2\pi $. This agrees
with the result obtained in Ref.~\cite{chm} by expressing $C(T)/T$
via the self-energy.

Consider now what happens when we add higher-order terms in $U$. They lead
to two types of corrections: self-energy corrections to the fermionic lines
in the two bubbles and corrections to the four-fermion vertices. The
self-energy corrections are of the FL type: they account for the appearance
of the quasiparticle $Z$ -factors and for the replacement of the bare
fermionic mass $m$ by $m^{\ast }$. Vertex corrections generate terms with
more bubbles. A generic diagram of $n$-th order has $n$ bubbles. To obtain a
$T^{3}$-contribution to $\Xi (T)$, we need to take dynamic, $\Omega _{m}/Q$
-terms from two bubbles out of $n$ and set $\Omega _{m}=0,Q\rightarrow 0$ in
the rest $n-2$ bubbles, because any extra power of $\Omega _{m}/Q$
eliminates the logarithmic singularity in the frequency integrand in $\delta
\Xi (T)$. It is intuitively plausible that once two dynamic bubbles are
chosen at the $n-$th order, the rest of the $n-$th order diagram constitutes
the $n-$th order correction to the static four-point vertex. If this
conjecture is true, the series of the diagrams for the non-analytic $T^{3}-$
term in the thermodynamic potential can be re-expressed in terms of the
 two-bubble diagrams in which  $U(0)$ and $U(2k_{F})$ are replaced
by exact \textit{static} vertices $\Gamma (\mathbf{k,-k;k,-k})$ and $\Gamma (%
\mathbf{k,-k;-k,k})$.
 Accordingly, $\Delta C(T)/T$ is given by
the same expression as in (\ref{genu}), but with $\Gamma $ instead of $U$.

This conjecture, however,
 needs to be verified as different diagrams for the
thermodynamic potentials contain different combinatorial factors, and it is
\textit{a priori} unclear whether these factors, combined with those counting
the number of ways two dynamic bubbles can be chosen, give the right
coefficients in the perturbative series for the static vertices. To verify
that this is the case, we evaluated explicitly the $T^{3}$-term in the
thermodynamic potential to third order in $U(q)$, and compared the result
with that given by the two-bubble diagrams with the renormalized static
vertices, evaluated independently. We found that the two expressions are
 identical. In what follows, we assume that this equivalence survives
to all orders in $U(q)$.

The renormalization from $U(0)$ and $U(2k_F)$ to $\Gamma (\mathbf{k,-k;k,-k})
$ and $\Gamma (\mathbf{k,-k;-k,k})$  includes static corrections coming from
the states both away and near the Fermi surface (the latter produce powers
of static $\Pi (\Omega _{m}=0,Q\rightarrow 0)=-m/2\pi $~\cite{agd}). In
other words, $\Gamma (\mathbf{k,-k;k,-k})$ and $\Gamma (\mathbf{k,-k;-k,k})$
include all vertex corrections except for the terms coming from the dynamic
part of the polarization bubble. In conventional notations~\cite{agd}, the
fully renormalized $\Gamma (\mathbf{k,-k;k,-k})$ and $\Gamma (\mathbf{%
k,-k;-k,k})$ are then related to $\Gamma ^{k}(\theta =\pi )$, which is the
limit of $\Omega =0$ and $q\rightarrow 0$ of $\Gamma (\mathbf{k},\mathbf{p},%
\mathbf{k+q},\mathbf{p-q})$, where both $\mathbf{k}$ and $\mathbf{p}$ are on
the Fermi surface, and $\theta $ is the angle between these two vectors.

The matrix (in the spin space) $\hat{\Gamma}^{k}(\pi )$ is related to the
quasiparticle scattering amplitude~ via $\hat{A}(\pi )=Z^{2}\hat{\Gamma}%
^{k}(\pi )$ \cite{agd}. Decomposing the scattering amplitude
for a spin-invariant interaction
 into charge and spin components $A_{c}$ and $A_{s}$ as
\begin{equation}
\hat{A}(\pi )=\frac{\pi v_{F}^{\ast }}{k_{F}}\left[ A_{c}(\pi )\hat{I}%
+A_{s}(\pi )\mathbf{\hat{\sigma}}\cdot
\mathbf{\hat{\sigma}}\right] \label{jul3_1}
\end{equation}
and comparing (\ref{jul3_1}) with a similar decomposition for $\hat{\Gamma}%
^{k}(\pi )$
\begin{eqnarray}
\hat{\Gamma}^{k}(\pi ) &=&\Gamma ^{k}(\mathbf{k,-k;k,-k})\hat{I}  \notag \\
&&-(1/2)\Gamma ^{k}(\mathbf{k,-k;-k,k})(\hat{I}+\mathbf{\hat{\sigma}}\cdot
\mathbf{\hat{\sigma}}),
\end{eqnarray}
we obtain
\begin{eqnarray}
Z^{2}\Gamma ^{k}(\mathbf{k,-k,k,-k}) &=&\frac{\pi v_{F}^{\ast }}{k_{F}}\left[
A_{c}(\pi )-A_{s}(\pi )\right] ;  \notag \\
Z^{2}\Gamma ^{k}(\mathbf{k,-k;-k,k}) &=&-2\frac{\pi v_{F}^{\ast }}{k_{F}}%
A_{s}(\pi ).  \label{jul3_3}
\end{eqnarray}
Substituting $Z^{2}\Gamma ^{k}(\mathbf{k,-k,k,-k})$ instead of $U(0)$ and $%
Z^{2}\Gamma ^{k}(\mathbf{k,-k;-k,k})$ instead of $U(2k_{F})$ into (\ref{genu}%
), we find the non-analytic part of $C(T)$ in a generic Fermi liquid
\begin{equation}
\Delta C(T)/T=-\left( ~\frac{3\zeta (3)}{2\pi }\right) ~\left(
\frac{m^{\ast }}{k_{F}}\right) ^{2}~~\left[ A_{c}^{2}(\pi
)+3A_{s}^{2}(\pi )\right] T. \label{jul3_4}
\end{equation}

A similar generalization of the second-order result for $\chi _{s}(T)$
\cite{chm}
yields
\begin{equation}
\Delta \chi _{s}\left( T\right) =\frac{m^{\ast }}{4k_{F}^{2}}\chi
_{s}\left( 0\right) A_{s}^{2}\left( \pi \right) T,  \label{chi}
\end{equation}
where $\chi _{s}\left( 0\right) $ is the spin susceptibility at $T=0$.

Eqs. (\ref{jul3_4}) and (\ref{chi}) are the two main results of
the paper. We see that the non-analytic parts in $C(T)$ and $\chi
_{s}(T)$ are parameterized by two scattering amplitudes,
$A_{c}(\pi )$ and $A_{s}(\pi ).$ These amplitudes are the new
parameters in an extended version of the FL
theory, which includes non-analytic terms. Notice that $A_{c}(\pi )$ and $%
A_{s}(\pi )$ cannot be simply expressed in terms of the
quasiparticle
interaction function $F (\theta )$, which, we remind, is related to $%
Z^{2}\hat{\Gamma}^{\omega }(\theta )=\left( \pi v_{F}^{\ast }/k_{F}\right)
\left( F _{c}(\theta )\hat{I}+F _{s}(\theta )\mathbf{\hat{\sigma}}%
\cdot \mathbf{\hat{\sigma}}\right) $.
 The harmonics $A_{a}^{(n)}$ and $F _{a}^{(n)}$ ($a=c,s$) are indeed
simply related, $A_{a}^{(n)}=F _{a}^{(n)}/(1+F _{a}^{(n)})$ in 2D, but
$A_{a}(\pi )$ is only expressed via infinite series of harmonics of
the Landau function:
\begin{equation}
A_{a}\left( \pi \right) =\sum_{n=0}^{\infty }(-1)^{n}F
_{a}^{\left( n\right) }/(1+F _{a}^{\left( n\right) }).
\label{fvsgamma}
\end{equation}
Rather involved explicit relations between $A_{a}(\pi )$ and $F
_{a}(\pi )$ to third order in $U$ are presented in
~\cite{unpublished}.

We find, however, that a simple relation between $\hat{A}(\pi )$ and $\hat{%
F}(\pi )$ does exist if the interaction $U(q)$ is strongly peaked at $%
q=0$. In this situation, only corrections due to $U(0)$ matter.
These corrections come from ring diagrams and can be summed up
exactly [the combinatorial factor for a ring diagram of order $n$
is $(-1)^{n}2^{n-2}(n-1) $]. Evaluating the sum over $n$, we
obtain $A_{c}(\pi )=F _{c}(\pi )/(1+F _{c}(\pi ))$. If, in
addition, $F _{s}(\theta )\ll 1$, the contribution to the specific
heat from $F _{c}(\pi )$ dominates, and the singular term in the
specific heat becomes
\begin{equation}
\Delta C\left( T\right) /T=-~\frac{3m\zeta (3)}{4\pi }~\frac{T}{E_{F}}%
~\left( \frac{F _{c}(\pi )}{1+F _{c}(\pi )}\right) ^{2}.
\label{genu_111}
\end{equation}
If $F_c (\pi)$ is dominated by the $n=0$ harmonic,
  Eq.~(\ref{genu_111}) coincides with the result of Ref.~\cite{aleiner}.

The limit $F _{c}(\pi )\rightarrow \infty $, $F _{s}(\theta )\ll
1$
corresponds to the Coulomb interaction for small $r_{s}$. In this limit, $%
F _{c}(\pi )$ is canceled out from (\ref{genu_111}), and the
singular term in the specific heat becomes
\begin{equation}
\Delta C\left( T\right) /T=-~\frac{3m\zeta (3)}{4\pi }~\frac{T}{E_{F}}.
\label{genu_1111}
\end{equation}
In agreement with Refs.~\cite{aleiner,dassarma}, we find that for the
Coulomb interaction, the $T^{2}$ term in the specific heat is independent of
$r_{s}$ for small $r_{s}$.

Scattering amplitudes in Eqs. (\ref{jul3_4}) and (\ref{chi}) can be
extracted from a measurement of $C(T)/T$ and $\chi _{s}\left( T\right) $ on
the same system. To the best of our knowledge, a linear-in-$T$ dependence of
$\chi_s$ has not been measured yet. However, the linear temperature dependence of $C(T)/T$
has been observed in several experiments on fluid monolayers He$^{3}$
adsorbed on graphite~\cite{greywall_2,ocura,casey}.
 To a reasonable accuracy, the
data can be fitted into a form $%
C/(NT/E_{F}^{\ast })=\gamma (T/E_{F}^{\ast })$, where $N$ is the density
per unit
area in a fluid monolayer, $E_{F}^{\ast }=E_{F}(m/m^{\ast })$ and $%
\gamma (x)\approx a-bx$ for small $x$ \cite{casey}.
Both $a$ and $b$ vary somewhat with $N$, but the variation is not dramatic, and to reasonable accuracy $%
a\sim 3-3.3$, and $b\sim 10-14$~ \cite{comm_a}. According to
 Eqs.(\ref{e2}) and (\ref{jul3_4}%
), $a=\pi ^{2}/3\approx 3.3\mathbf{\ }$, and $b=0.9[A_{c}^{2}\left(
  \pi
\right) +3A_{s}^{2}\left( \pi \right) ]$. A fit to the data then yields $%
A_{c}^{2}\left( \pi \right) +3A_{s}^{2}\left( \pi \right) \approx
11-15.5$~ \cite{comm_a}. To estimate $A_{c}\left( \pi \right) $ and
$A_{s}\left( \pi \right) $ separately, we assume that the scenario of
``almost localized fermions'' \cite{voll}, which describes
successfully the properties of bulk He$^{3},$ is applicable to the 2D
case as well. In this scenario, the interaction in the charge channel
is strong, whereas that in the spin channel is moderate.  A strong
interaction in the charge channel means that $F _{c}^{\left( n\right)
  }\gg 1$, in which case the consecutive terms in series for
$A_{c}\left( \pi \right) $ [Eq.(\ref{fvsgamma})] almost cancel
each other, and the result is likely to be small. A precsie value for $%
A_{c}\left( \pi \right) $ depends on how $F _{c}^{\left( n\right)
}$ decrease with $n.$ However, in two model cases $F _{c}^{\left(
n\right) }=g/(1+n^{2})$ and $F _{c}^{\left( n\right) }=ge^{-n}$,
we obtain almost identical results: $A_{c}^{2}\left( \pi \right)
\approx 0.25$ in the limit of $g\gg 1.$ This suggests that the
observed value $A_{c}^{2}\left( \pi \right) +3A_{s}^{2}\left( \pi
\right) \approx 11-15.5$ is almost entirely due to the spin part
of the amplitude. Neglecting $A_{c}^{2}\left( \pi \right) ,$ we
obtain $|A_{s}\left( \pi \right) |\approx 1.\,\allowbreak 9-2.3.$
If the $n=0$ harmonic of $F _{s}$ dominates the result for
$A_{s}\left( \pi \right)$, \emph{i.e.}, $A_s \left( \pi \right)
\approx F _{s}^{(0)}/(1+F _{s}^{(0)})$, then $F _{s}^{(0)}\approx
-(0.66-0.7)$, which is consistent with the 3D
 value $F _{s}^{\left( 0\right) }\approx -0.75$%
\emph{~}\cite{greywall}.

Notice also that if $A_{c}(\pi)$ can be neglected compared to
$A_s(\pi)$, $\Delta C(T)/T$ and $\delta \chi _{s}(T)$ contain only one unknown
parameter ($A_s(\pi)$).
 In this situation, the ratio $K=\Delta C(T)/(T\Delta \chi _{s}(T))$ is
expressed only via the parameters describing the leading, analytic parts of $%
C(T)$ and $\chi _{s}$: $K=-18\zeta (3)m^{\ast }/\pi \chi _{s}(0)$.

In addition, $A_s(\pi)$ determines the slope of the linear-in-$T$
correction $\Delta\sigma (T)$ to the conductivity of a dirty 2D FL in
the ballistic 
regime \cite {zna}, which allows to express $\delta\chi_s (T)$ in
terms of $\Delta\sigma(T)$ as
$\Delta\chi_s(T)E_F^*/\chi_s(0)T=\left[1-\Delta\sigma (T)
E_F^*/\sigma(0)T\right]^2/72$. These predictions are amenable to a
direct experimental verification.

To summarize, in this paper we showed that the 2D Fermi-liquid
theory describes not only the leading, constant terms in the
specific heat coefficient $C(T)/T$ and the spin susceptibility
$\chi _{s}(T)$, but also subleading, linear-in-$T$ terms. We
argued that these terms come from backscattering, and are
universally expressed via the spin and charge components of the
scattering amplitude at the angle $\theta =\pi $. We extracted the
spin component of the scattering amplitude from the experimental
data on $C\left( T\right) $ for a monolayer of He$^{3}.$

We acknowledge stimulating discussions with I. Aleiner, B. Altshuler, A.
Andreev, A. Millis, and J. Saunders. The research has been supported by NSF
Grant No. DMR 0240238 (A. V. Ch.), NSF Grant No. DMR-0308377 (D. L. M.), and
NSF Grant No. DMR-0237296 (L. I. G.).

\end{document}